\documentclass[aip,rsi, amsmath,amssymb, reprint, fontsize=12pt]{revtex4-1}
\pdfoutput=1
\usepackage{graphicx}
\usepackage{dcolumn}
\usepackage{bm}

\usepackage{pgfplots}
\usepackage{pgfplotstable}
\usepgfplotslibrary{external} 
\usepgfplotslibrary{fillbetween}
\usepgfplotslibrary{colorbrewer}
\pgfplotsset{compat=1.6}
\usepackage{lineno}

\usetikzlibrary{external}
\usepackage{subfigure}
\usepackage{cleveref}

\pgfmathsetmacro{\Tfreeze}{364}
\pgfmathsetmacro{\Tjump}{442.335666891}
\pgfmathsetmacro{\tfreeze}{0.483233608052}
\pgfmathsetmacro{\tjump}{0.535762818744}
\pgfmathsetmacro{\Tc}{537}

\definecolor{android_blue}{RGB}{51,181,229}
\definecolor{android_dark_blue}{RGB}{0,153,204}
\definecolor{android_pink}{RGB}{170,102,204}
\definecolor{android_purple}{RGB}{156,39,176}
\definecolor{android_dark_pink}{RGB}{153,51,204}
\definecolor{android_green}{RGB}{153,204,0}
\definecolor{android_dark_green}{RGB}{102,153,0}
\definecolor{android_orange}{RGB}{255,152,0}
\definecolor{android_dark_orange}{RGB}{255,152,0}
\definecolor{android_red}{RGB}{255,68,68}
\definecolor{android_dark_red}{RGB}{204,0,0}
\definecolor{android_pink}{RGB}{156,39,176}
\definecolor{android_grey}{RGB}{158,158,158}

\definecolor{olivia_red}{RGB}{215,25,28}
\definecolor{olivia_orange}{RGB}{253,174,97}
\definecolor{olivia_yellow}{RGB}{255,255,191}
\definecolor{olivia_lightblue}{RGB}{171,217,233}
\definecolor{olivia_blue}{RGB}{44,123,182}

\pgfplotsset{grid style={dashed,grey,opacity=0.5}}

\pgfplotscreateplotcyclelist{peak_temp}{
{color=android_dark_blue,line width=1.0pt,mark=x,mark size=2pt,mark options={line width=1.0pt},line join=round},
{color=android_red,line width=1.0pt,mark=o,mark size=2pt,mark options={line width=1.0pt},line join=round},
{color=android_dark_green,line width=0.5pt,mark=|,mark size=2pt,mark options={line width=0.75pt},line join=round},
{color=black,line width=0.75pt,mark size=2pt,mark=triangle,mark options={line width=0.75pt},line join=round},
{color=android_blue,line width=0.5pt,mark=square,mark size=2pt,mark options={line width=0.75pt},line join=round},
	  {color=android_pink,line width=0.5pt,mark=diamond,mark size=2pt,mark options={line width=0.75pt},line join=round},
	  {color=android_orange,line width=0.5pt,mark size=2pt,mark options={line width=0.75pt},line join=round}
	  }

\begin{document}
\pgfplotsset{colormap/RdBu-9}

\title{Write head design for effective curvature reduction in heat-assisted magnetic recording by topology optimization }

\author{O. Muthsam}
 \email{olivia.muthsam@univie.ac.at}
\author{C. Vogler}
\author{F. Bruckner}
\author{D. Suess}
\affiliation{ 
University of Vienna, Physics of Functional Materials, Boltzmanngasse 5, 1090 Vienna, Austria
}%

\date{\today}
             
\begin{abstract}
The reduction of the transition curvature of written bits in heat-assisted magnetic recording (HAMR) is expected to play an important role for the future areal density increase of hard disk drives. Recently a write head design with flipped write and return poles was proposed. In this design a large spatial field gradient of the write head was the key to significantly reduce the transition curvature.  
In this work we optimized the write pole of a heat-assisted magnetic recording head in order to produce large field gradients as well as large fields in the region of the heat pulse. This is done by topology optimization. The simulations are performed with dolfin-adjoint. For the maximum field gradients of 8.1\,mT/nm, 8.6\,mT/nm and 11.8\,mT/nm, locally resolved footprints of an FePt like hard magnetic recording medium are computed with a coarse-grained Landau-Lifshitz-Bloch (LLB) model and the resulting transition curvature is analysed. Additional simulations with a bilayer structure with $50\%$ hard and $50\%$ soft magnetic material are computed. The results show that for both recording media, the optimized head design does not lead to any significant improvement of the written track. Thus, we analyse the transition curvature for the optimized write heads theoretically with an effective recording time window (ERTW) model. Moreover, we check how higher field gradients influence the curvature reduction. The results show that a simple optimization of the conventional head design design is not sufficient for effective curvature reduction. Instead, new head concepts will be needed to reduce transition curvature. 

\end{abstract}

\maketitle

\section{Introduction}
\begin{figure}
\centering
\includegraphics[width=1.0\linewidth]{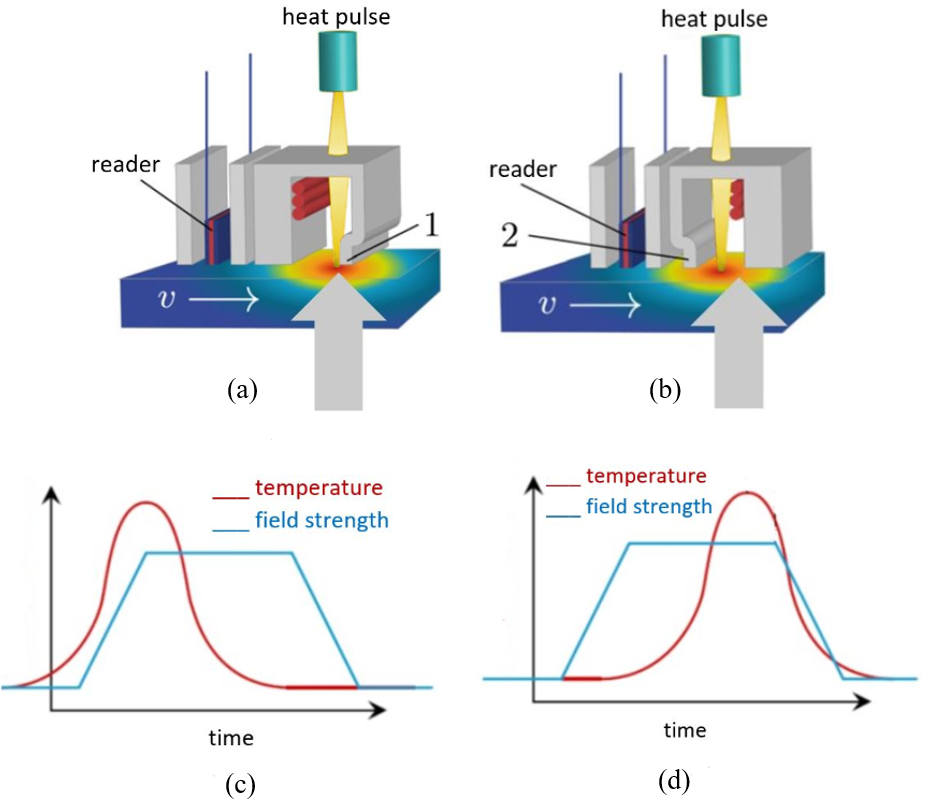}
\caption{Schematic representation of (a) a conventional and (b) a flipped recording head design proposed by Vogler \textit{et al} \cite{vogler_reduction}. In (c) and (d) the temporal evolution of the applied field and the temperature pulse for the conventional and the flipped design, respectively, can be seen. The grey arrows in (a) and (b) indicate the position at which the grains are written.}
\label{proposeddesign}
\end{figure}

In heat-assisted magnetic recording (HAMR) \cite{ersteshamr,fan,hamr1,hamr} a heat pulse is included to the writing process to overcome the so-called recording trilemma \cite{evans} and make high-anisotropy grains writable with the available head fields. 
However, in granular media, the curved thermal profile of the heat pulse in combination with a spatially relatively homogeneous head field gives rise to a significant curvature at transitions between bits \cite{nft1,nft2}. This transition curvature is expected to be a serious problem for the read-back process in HAMR since the signal-to-noise ratio (SNR) is reduced \cite{zhu_correcting}. Different methods to efficiently reduce transition curvature in HAMR have been proposed, for example a write head field design by Zhu \textit{et al} \cite{zhu_field,zhu_correcting,zhuli}. The present work is based on the publication by Vogler \textit{et al} \cite{vogler_reduction}, where a recording head design with flipped write and return pole to efficiently reduce transition curvature is suggested.
For the flipped head design writing of the bit happens between the near field transducer (NFT) and the returnpole, whereas for the conventional design writing happens near the write pole. The position where the bit is written for the different design is indicated by the grey arrows in \Cref{proposeddesign}(a) and (b).
The different writing position leads to a different behavior of the write field during the cooling process of the heat pulse. For the conventional design, the applied field is spatially relatively homogeneous during the cooling process (see \Cref{proposeddesign}(c)). In contrast to this, shown in \Cref{proposeddesign}(d), the field decreases during cooling for the flipped design which leads to a field gradient in down-track direction. This field gradient turned out to be the key for the curvature reduction. 
For this reason, the head field gradient in down-track direction is an important parameter in the simulations. \\
In this work, we compute the realistic field gradient for a flipped design, that follows the model of a state-of-the-art recording head design \cite{suh_head, lee_head}, at the position where the heat pulse is cooling down. To optimize the head, we design a write pole that maximizes both the field and the field gradient at the required position. This is done by topology optimization \cite{bendsoe_topology_2013}, an application of the inverse magnetostatic problem \cite{bruckner_solving}. With the resulting fields and field gradients we compute locally resolved switching probability phase diagrams with a coarse-grained LLB model \cite{LLB} for both pure hard magnetic recording material and an exchange spring structure with $50\%$ hard and $50\%$ soft magnetic material. From this locally resolved phase diagrams, we can then analyse the transition curvature.\\
    Additionally, we interpret the results in the context of the effective recording time window (ERTW) model by Vogler \textit{et al} \cite{basic,vogler_reduction}.
This paper is structured as follows: In Section II, a theoretical framework of the simulation methods is presented and the simulation and material parameters are given. The results are presented in Section III and discussed in Section IV.

\section{Theoretical framework}

\subsection{Topology optimization}
To solve the topology-optimization problem, the density method, which is also known as solid isotropic microstructure with penalization (SIMP), is used \cite{huber_topology,abert2017fast}. First, the geometry is meshed with a tetrahedral mesh. At each element of the mesh the density $\rho$ of the material is considered and a value between 0 (void) and 1 (material) is assigned to it. This leads to one optimization variable per element. The final design should only have density values of 0 or 1. This is achieved by using a penalization parameter $k$ to penalize those densities that are intermediate. \\
The materials in the simulations are approximately described by linear material laws.
For the soft magnetic material, the material law reads as

\begin{align}
\textbf{B}_{\textbf{s}} = \mu_0 (\textbf{H}_{\textbf{s}} + \textbf{M}_{\textbf{s}}) = \mu_0 \mu_r \textbf{H}_{\textbf{s}}
\end{align}

where $\mu_r$ is the relative permeability. $\mu_0$ denotes the vacuum permeability.
For the hard magnetic material, there holds

\begin{align}
\textbf{B}_{\textbf{p}} = \mu_0 \mu_m \textbf{H}_{\textbf{p}} + \mu_0 \textbf{M}_{\textbf{0}}
\end{align}

with the recoil permeability $\mu_m$. For a permanent magnetic region $\Omega_{\mathrm{p}} \subset \mathbb{R}^3$, the magnetization as a function of the density $\rho$ can be written as

\begin{align}
\textbf{M}(\rho)=\rho^k\textbf{M}_0
\end{align}
with the density value $\rho \in [0,1]$ of one element and the penalization parameter $k$. For hard magnetic materials, $k=1$ is a good choice \cite{choi}.
For a soft magnetic design region $\Omega_{\mathrm{s}}\subset \mathbb{R}^3$, the magnetic susceptibiliy $\chi$ can be reformulated as a function of the density to

\begin{align}
\chi(\rho)=\chi_0 \cdot \rho^k
\end{align}

and thus for the relative permeability, there holds

\begin{align}
\mu_r(\rho)=(\mu_{r0}-1)\rho^k + 1.
\end{align}

In this way it can be used for topology optimization. In the case of soft magnetic material, $k=4$ leads to good results \cite{huberthesis}.\\
The topology-optimization problem that needs to be solved is given by

\begin{equation}
\begin{aligned}
\mathrm{Find:} \min_{\rho} J(\rho) \\
\mathrm{subject\, to:} \int_{\Omega_i} \rho(r) dr \leq V;\\
0 \leq \rho(r) \leq 1, r \in \Omega_i
\end{aligned}
\end{equation}

where $J$ is the objective function, $V$ is the maximum volume of the design as a constraint and $i \in \{s,p\}$ defines the soft and permanent magnetic region, respectively.\\
The objective function used to maximize the magnetic field and the $z-$field gradient is given by minimizing

\begin{equation}
\begin{aligned}
J(\rho) = \int_{\Omega_{\mathrm{field}}} \frac{1}{|\nabla H_z(\rho)|^2} dr +\, \int_{\Omega_{\mathrm{field}}} \frac{1}{|H_z(\rho)|^2} dr \\ + \,\underbrace{\lambda \cdot \int_{\Omega_{\mathrm{opt}}} \rho(1-\rho) dr}_{\mathrm{penalization}},
\end{aligned}
\end{equation}

where $\Omega_{\mathrm{opt}}$ is the region of the material that is optimized, $H_z(\rho)$ is the $z-$field and $\nabla H_z(\rho)$ is the gradient of the $z-$field. $\lambda$ is a penalization parameter. 
The topology-optimization problem is solved by a finite element method (FEM) which is based on the open-source library FEniCS \cite{fenics1,fenics2} for solving partial differential equations (PDEs) and the library dolfin-adjoint\cite{dolfin1,dolfin2}. Dolfin-adjoint automatically determines and solves adjoint linear equations using PDEs which are discretized with finite elements. The minimization problem is solved using the L-BFGS-B method \cite{lbfgsb1,lbfgsb2}, a limited-memory quasi-Newton solver for bound-constrained optimization.

\subsection{Head design parameters}
\begin{figure}
\centering
\includegraphics[width=1.0\linewidth]{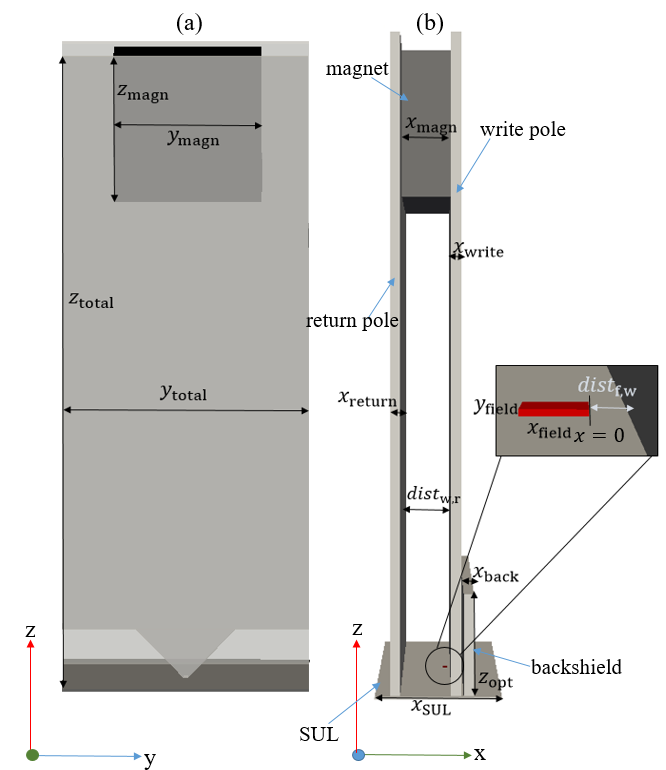}
\caption{Schematic representation with dimensions of the initial flipped head design. The dimensions can be seen in \Cref{tabledimensions}. In (a) the front view and in (b) the side view can be seen. The 100\,nm $\times$ 100\,nm wide fieldbox is 50\,nm away from the write pole.}
\label{backshield100}
\end{figure}

In the topology-optimization simulations, a write head \cite{recording_head,brug_magnetic_1996} is optimized, which consists of a write and a return pole, a permanent magnet to simulate the core magnetized by a coil and a 50\,nm thick soft magnetic underlayer (SUL). Additionally, in some simulations a backshield is considered. All components and the dimensions of the write head are marked in \Cref{backshield100} and summarized in \Cref{tabledimensions}. The tip of the write pole and the backshield, if considered, are the parts of the write head that are optimized by topology optimization. Except the permanent magnet, all other parts of the write head consist of soft magnetic material.\\
The material of the recording head is assumed to be FeCo \cite{kief_materials}. Hence, a relative permeability $\mu_{\mathrm{r}}=18000$ and a saturation polarization of 2.4\,T are assumed \cite{feco}.
It is considered that the coil magnetizes the core with 0.8\,T in $x-$direction which is modeled by the means of a permanent magnet that is magnetized in $x-$direction with 0.8\,T. This field is chosen because then the maximum magnetization of the initial design is slightly below the saturation polarization of the material.\\
The field and the field gradient are computed and optimized at the position where the recording medium hitted by the heat spot produced by the NFT is already cooling down. This position is assumed to be 50\,nm away from the pole tip in $x-$direction.\\
Since small head to media spacings (HMS) are needed, to get high areal storage density \cite{hmsmedia}, the HMS is assumed to be 5\,nm.\\
Four different starting geometries are considered. The starting geometries are basic geometries of dimensions $x_{\mathrm{write}}\times y_{\mathrm{total}} \times z_{\mathrm{opt}}$, where the density initially is 1 for each element. During the optimization process, the density of each element is adjusted. The difference between the geometries is the dimension in $x-$direction and the fact if a backshield is considered or not. The dimensions in $y-$ and $z-$direction are the same for all starting geometries and equal to $y_{\mathrm{total}}$ and $z_{\mathrm{opt}}$, respectively. The first geometry to be optimized, is one with dimension $x_{\mathrm{write}}=100$\,nm and no backshield. This one is referred to as Basic$_{100}$. The second geometry also has dimension $x_{\mathrm{write}}=100\,$nm. However, for this geometry a backshield with $x_{\mathrm{back}}=100\,$nm is additionally considered and optimized. Afterwards, this geometry is called Backshield$_{100}$. The last geometries are similar to the former ones but with dimensions $x_{\mathrm{write}}=200$\,nm and $x_{\mathrm{back}}=200$\,nm. They are labeled Basic$_{200}$ and  Backshield$_{200}$. The $x_{\mathrm{write}}$ and $x_{\mathrm{back}}$ values of the different geometries are summarized in \Cref{tablegradients}.\\
After the optimal head designs are determined via topology-optimization, additional simulations with magnum.fe \cite{abert_magnum.fe:_2013} that include a coil instead of a permanent magnet are performed. These coil-simulations are performed in order to determine realistic fields with the optimized head designs. Here, all parts of the write head are considered to be soft magnetic with the above material parameters. The current density inside the coil is assumed to be $2.5\times 10^{10}$\,A/m$^2$, which is below the current density limit of $10^{12}\,$A/m$^2$, \cite{hamann2007heating}. 

\section{Results}

\begin{center}
\begin{table*}
\centering
\begin{tabular}{|>{\centering}m{1.1cm}|>{\centering}m{1.1cm}|>{\centering}m{1.1cm}|>{\centering}m{1.1cm}|>{\centering}m{1.1cm}|>{\centering}m{1.1cm}|>{\centering}m{1.2cm}|>{\centering}m{1.2cm}|>{\centering}m{1.2cm}|>{\centering}m{1.2cm}|c|c|c|c|c}\hline
  $x_{\mathrm{SUL}}$  & $y_{\mathrm{total}}$ &$z_{\mathrm{total}}$& $x_{\mathrm{magn}}$& $y_{\mathrm{magn}}$& $z_{\mathrm{magn}}$& $z_{\mathrm{opt}}$&$x_{\mathrm{write}}$&$x_{\mathrm{return}}$& dist$_{\mathrm{w,r}}$& $x_{\mathrm{back}}$& dist$_{\mathrm{f,w}}$&$x_{\mathrm{field}}$ & $y_{\mathrm{field}}$ \\
    \hline
	2\,$\mu$m & 5\,$\mu$m & 13\,$\mu$m& 1\,$\mu$m&3\,$\mu$m&3\,$\mu$m& 2\,$\mu$m& varied &200\,nm & 1\,$\mu$m & varied&50\,nm&100\,nm&100\,nm\\
    \hline
 \end{tabular}
\caption{Approximated dimensions of an initial flipped recording head design as marked in \Cref{backshield100}. }
\label{tabledimensions}
\end{table*}
\end{center}

\subsection{Field gradients}

\begin{figure}
\dimendef\prevdepth=0
\centering
\includegraphics[width=1.0\linewidth]{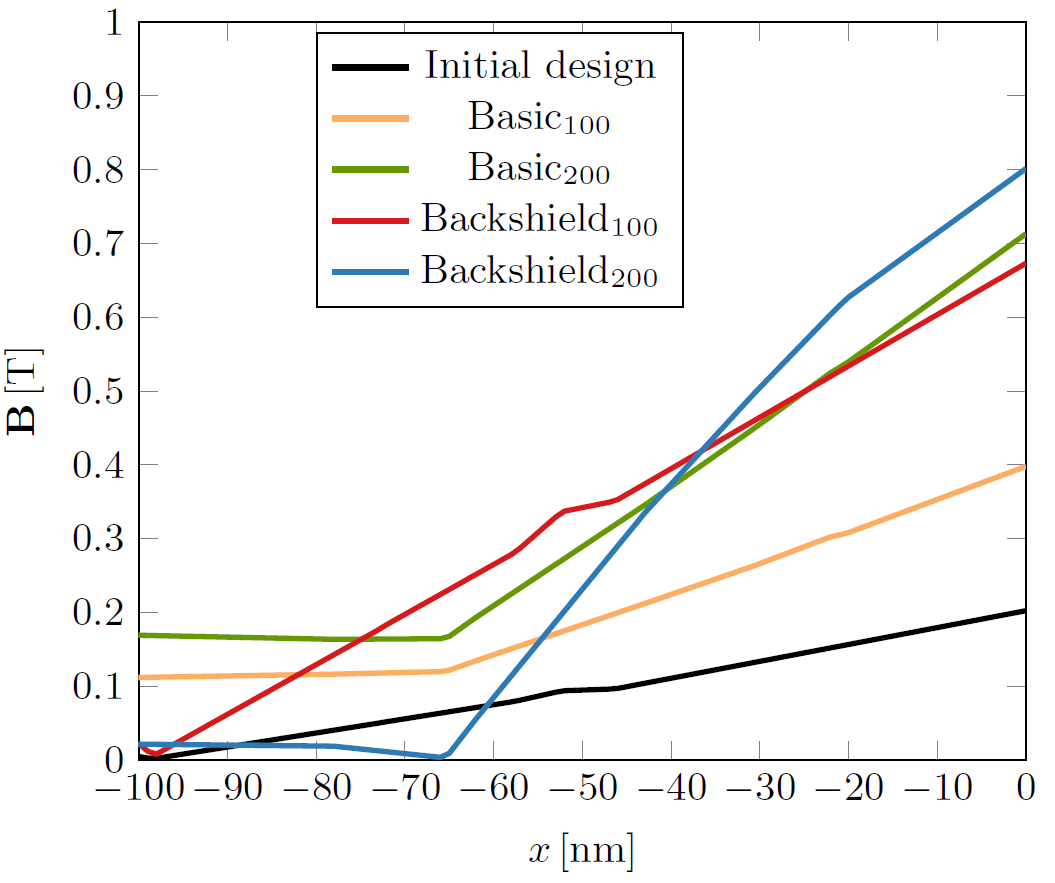}
\caption{Comparison of the fields of the topology optimized geometries for a FeCo like head material. The magnetic field is plotted over the $x-$length of the 100\,nm $\times$ 100\,nm wide fieldbox (marked red in \Cref{backshield100}). At $x=0$, the fieldbox is 50\,nm away from the edge of the write pole.}
\label{gradients}
\end{figure}

\begin{figure}
\centering
\includegraphics[width=1.0\linewidth]{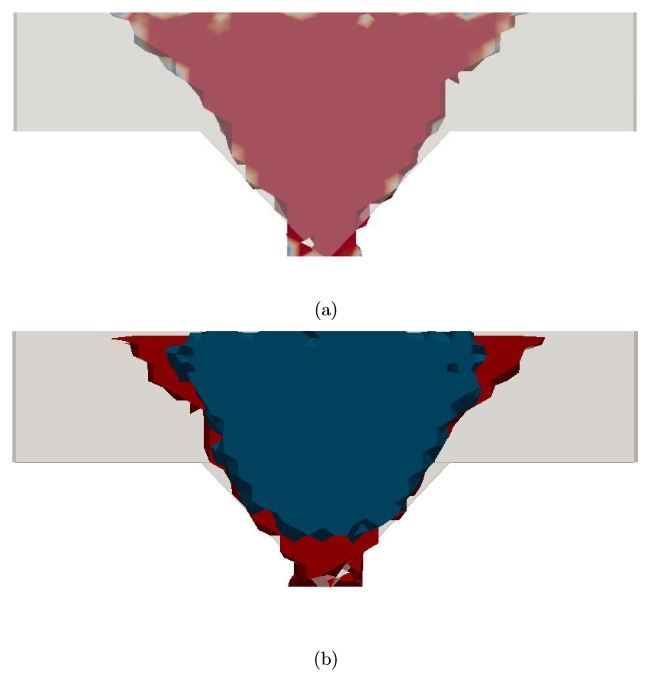}
\caption{Comparison of the initial flipped head design (grey) and the topology optimized write pole (red) with a backshield (blue) with $x_{\mathrm{write}}=200$\,nm and $x_{\mathrm{back}}=200$\,nm (red). (a) Front view and (b) rear view  with backshield of the optimized write pole.}
\label{recordinghead}
\end{figure}

Forward simulations show that the field gradient of the initial flipped design, which follows a state-of-the-art head design \cite{suh_head, lee_head}, at 50\,nm distance from the write pole is approximately 2.2\,mT/nm.
In \Cref{gradients} the resulting fields from the coil simulations with the optimized geometries are plotted. The results show that the best outcome can be achieved for a geometry with a pole tip with $x_{\mathrm{write}}=200$\,nm and an additional backshield with $x_{\mathrm{back}}=200$\,nm. Here, the maximum field gradient is 11.8\,mT/nm. The optimized Backshield$_{200}$ geometry is shown in \Cref{recordinghead}. Note that all optimized geometries are similar and show a tapered shape. Recapitulating, the resulting write fields and field gradients for the different geometries are summarized in \Cref{tablegradients}.

\begin{center}
\begin{table*}
\centering
\begin{tabular}{|l|c|c|c|c|}\hline
 Geometrie & $x_{\mathrm{write}}$\,[nm]& $x_{\mathrm{back}}$\,[nm]& $\mu_0H$\,[T] &Max. $\mu_0dH/dx$\,[mT/nm]\,\\
  \hline
  	Conventional&100&---&0.8&---\\
	Initial flipped design&100&---& 0.2 &  2.2\\
   Basic$_{\mathrm{100}}$ &100&---& 0.4 & 4.1\\
    Basic$_{\mathrm{200}}$& 200&---&0.7 &8.6\\
    Backshield$_{\mathrm{100}}$&100&100& 0.67 &8.1\\
    Backshield$_{\mathrm{200}}$&200&200& 0.8 &11.8\\
  \hline
 \end{tabular}
\caption{Resulting fields and field gradients produced by the write heads which are optimized with the help of topology optimization for a FeCo like head material. The field and its gradient of a conventional head are added for comparison.}
\label{tablegradients}
\end{table*}
\end{center}


\subsection{Curvature reduction}
To analyse the curvature reduction, switching probability phase diagrams are computed with the help of a coarse-grained model based on the stochastic Landau-Lifshitz-Bloch equation (LLB) \cite{LLB} for a FePt like granular recording medium. The material parameters of the medium can be seen in \Cref{tablematerialien}. In the simulations a continuous laser pulse with Gaussian shape and a full width at half maximum (FWHM) of 60\,nm is considered. The temperature profile of the heat pulse is given by

\begin{align}
T(x,y,t)= (T_{\mathrm{write}}-T_{\mathrm{min}})e^{-\frac{(x-vt)^2+y^2}{2\sigma^2}} + T_{\mathrm{min}} \\
= T_{\mathrm{peak}}(y)\cdot e^{-\frac{(x-vt)^2}{2\sigma^2}} + T_{\mathrm{min}}
\label{pulse}
\end{align}

with

\begin{align}
\sigma=\frac{\mathrm{FWHM}}{\sqrt{8\ln(2)}}.
\end{align}

\begin{figure}
\dimendef\prevdepth=0
\centering
\includegraphics[width=1.0\linewidth]{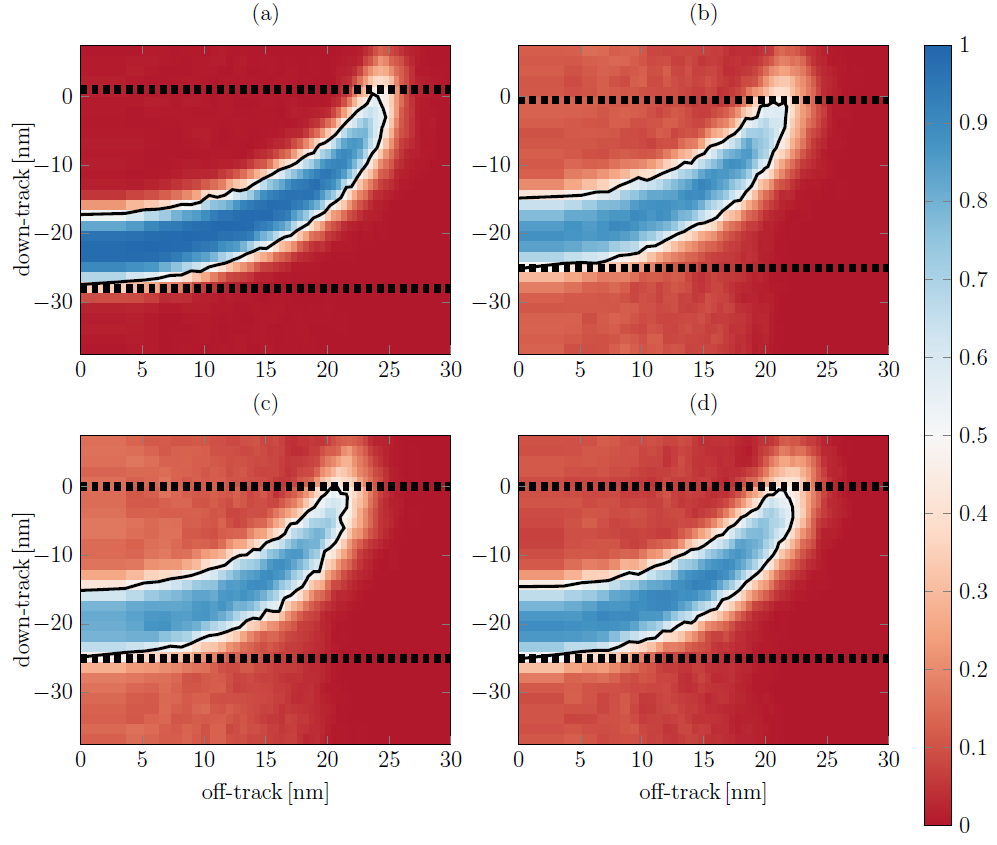}
  \caption{Switching probability phase diagrams of (a) a conventional head with a homogeneous write field and the flipped design with optimized field gradients (b) $\mu_0dH/dx$ = 8.6\,mT/nm (Basic$_{200}$), (c) $\mu_0dH/dx$ = 8.1\,mT/nm (Backshield$_{100}$) and (d) $\mu_0dH/dx$ = 11.8\,mT/nm (Backshield$_{200}$) in combination with pure hard magnetic recording material. The black dashed lines indicate the maximum and minimum down-track position $x$ with $P(x) = 50 \%$.} 
  \label{footprints}
\end{figure}

The speed $v$ of the write head is assumed to be 15\,m/s. $x$ and $y$ are the down-track and the off-track position of the grain, respectively. In the simulations both the down-track position $x$ and the off-track position $y$ are variable. The initial and final temperature of all simulations is $T_{\mathrm{min}}=300$\,K. The write temperature $T_{\mathrm{write}}$ is chosen to be $800\,$K. 
The applied field is modeled as a trapezoidal field with a field duration of 0.57\,ns and a field rise and fall time of 0.1\,ns. The angle of the applied field with respect to the normal is assumed to be 22\,deg. 
In each phase point 128 HAMR simulation simulations of a recording grain are performed.\\
In the phase diagrams the switching probability of a recording grain is shown as a function of the down-track position $x$ and the off-track position $y$. First, the phase diagram is computed for a spatially homogeneous field which approximates a conventional recording head design. Additionally, footprints for optimized geometries with field gradients of 8.6\,mT/nm (Basic$_{200}$), 8.1\,mT/nm (Backshield$_{100}$) and 11.8\,mT/nm (Backshield$_{200}$) are computed. The geometries with field gradients 2.2\,mT/nm (Initial flipped) and 4.1\,mT/nm (Basic$_{100}$) are not further considered since the phase diagrams show too much noise to get reliable results. Note, that different write fields (see \Cref{tablegradients}) are used for the simulations with different field gradients as seen in \Cref{tablegradients}. In \Cref{footprints}, the resulting switching probability phase diagrams can be seen. There are some visible differences between the footprints of the different field gradients. \\
One can see that for the flipped designs the recording performance is worse than for the conventional design where a homogeneous field is assumed. This can be seen by the increase of jitter in down-track direction and the reduction of the maximum switching probability $P_{\mathrm{max}}$. $P_{\mathrm{max}}$ and the jitter can be determined by fitting the $P(x)-$curve at one temperature with a Gaussian cumulative distribution function 

\begin{align}
\Phi_{\mu,\sigma^2}=\frac{1}{2} (1 + \mathrm{erf}(\frac{x-\mu}{\sqrt{2\sigma^2}}))\cdot p
\label{distribution}
\end{align}

with

\begin{align}
\mathrm{erf}(x)=\frac{2}{\sqrt{\pi}} \int_0^x e^{-\tau^2} d\tau,
\label{error}
\end{align}

where the fitting parameters are the mean value $\mu$, the standard deviation $\sigma$ and the mean maximum switching probability $p \in [0,1]$. The standard deviation $\sigma$ determines the steepness of the transition function and is a measure for the transition jitter. The temperature at which the down-track jitter is determined is $T_{\mathrm{peak}}=760$\,K. The resulting jitter and $P_{\mathrm{max}}$ values are summarized in \Cref{table}.
Additionally, it can be seen that the grains are written at larger off-track positions for increasing field gradients and the resulting lower write fields. 
%
%
The various write fields lead to writing of the grains at different peak temperatures. For smaller write fields the grains are written at higher temperatures only whereas they are written at lower temperatures for higher fields. The convention

\begin{align}
T_{\mathrm{peak}}(y)=(T_{\mathrm{write}}-T_{\mathrm{min}})e^{-\frac{y^2}{2\sigma^2}}+T_{\mathrm{min}}.
\label{equation}
\end{align}

shows that with higher write fields, the off-track edge of a bit shifts to larger off-track positions, because the minimum temperature necessary to write a grain is smaller. 

\begin{center}
\begin{table*}
\centering
\begin{tabular}{|>{\centering}m{2.2cm}|>{\centering}m{1.7cm}|>{\centering}m{3.0cm}|>{\centering}m{3.1cm}|>{\centering}m{1.5cm}|>{\centering}m{2.5cm}|c|c|}
\hline
Curie temp. $T_{\mathrm{\textbf{C}}}$ (K) & Damping $\alpha$& Anisotropy const. $K_{1}$ (MJ/m$^3$)& Anisotropy field $\mu_0H_k$ (T) at 300\,K&$J_{\mathrm{s}}$ (T)& height $h$ (nm) & diameter $d$ (nm) \\
    \hline
	693.5 & 0.02&$6.6$ & 10 & 1.35 &8 & 5 \\
    \hline
 \end{tabular}
\caption{Material parameters of a FePt like hard magnetic granular recording medium.  Note, that different material parameters for the recording medium were used compared to the original work of Vogler \textit{et al} \cite{vogler_reduction}. }
\label{tablematerialien}
\end{table*}
\end{center}


For a detailed analysis of the transition curvature, the full down-track range $\Delta x$ in which the bit is written with $P_{\mathrm{max}}\ge 50 \%$ is computed. $\Delta x$ is marked in \Cref{footprints} as the distance between the black dashed lines. Because of the different off-track widths $\Delta y$ in the phase diagrams for the different geometries, it is necessary to scale the curvature parameter with $\Delta y$. This is done, since the track width is usually kept constant in magnetic recording. In reality the track-width can for example be steered by controlling the peak temperature $T_{\mathrm{peak}}$ or the full width at half maximum \cite{huang2014heat}. With the curvature parameter $c=\Delta x/\Delta y$, the curvature can be reliably analyzed. Note that $c$ is equivalent to the curvature parameter used by Vogler \textit{et al} \cite{vogler_reduction} when multiplied by $\Delta y$ for scaling reasons. There holds

\begin{align}
    c=\frac{\Delta x}{\Delta y} \equiv cp\cdot \Delta y
\end{align}

where $cp$ is the curvature parameter defined by Vogler \textit{et al} \cite{vogler_reduction}.
The curvature analysis shows that the curvature is increased for both the Basic$_{200}$ and the Backshield$_{100}$ design compared to the conventional design. The curvature reduction for the Backshield$_{200}$ flipped head design with a field gradient of $11.8$\,mT/nm is about $1.3\%$. Detailed information about the curvature parameters can be seen in \Cref{table}.

\begin{center}
\begin{table*}
\centering
\begin{tabular}{|l|c|>{\centering}m{3.0cm}|c|r|}\hline
   Geometry & $c$ ($\%$) \,\,\,& $P_{\mathrm{max}}$ ($\%$)  &$\sigma_{\mathrm{down}}$ ($\%$) \\
    \hline
    Basic$_{200}$&$+1.4$&$-5.5$&$+118.7$\\
    Backshield$_{100}$&$+3.8$&$-7$&$+113.5$\\
    Backshield$_{200}$&$-1.35$&$-2.3$ & $+28.4$\\
  \hline
 \end{tabular}
\caption{Resulting transition curvature, $P_{\mathrm{max}}$ and jitter parameters of the different flipped head designs in combination with a pure hard magnetic recording material compared to a conventional recording head design.}
\label{table}
\end{table*}
\end{center}

\subsection{Exchange Spring Recording Medium}
\begin{figure}
\dimendef\prevdepth=0
\centering
\includegraphics[width=1.0\linewidth]{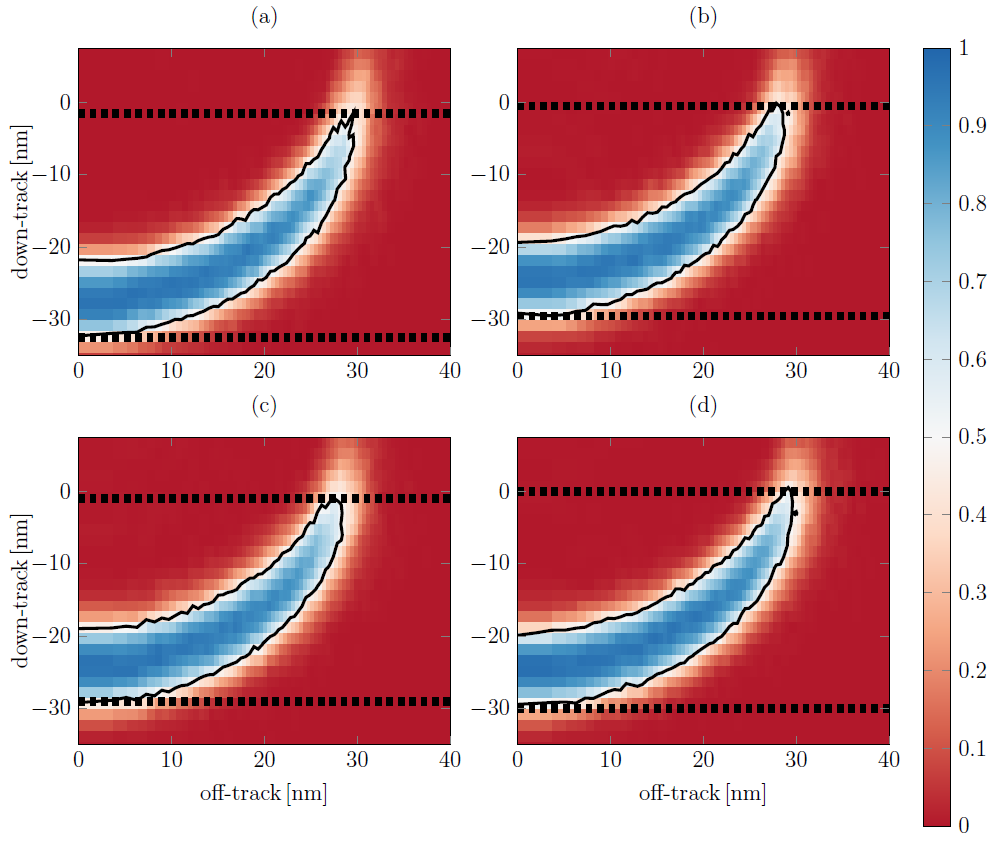}
  \caption{Switching probability phase diagrams of (a) a conventional head with a homogeneous write field and the flipped design with optimized field gradients (b) $\mu_0dH/dx$ = 8.6\,mT/nm (Basic$_{200}$), (c) $\mu_0dH/dx$ = 8.1\,mT/nm (Backshield$_{100}$) and (d) $\mu_0dH/dx$ = 11.8\,mT/nm (Backshield$_{200}$) in combination with an exchange spring bilayer structure. The black dashed lines indicate the maximum and minimum down-track position $x$ with $P(x) = 50 \%$.} 
  \label{footprintssm}
\end{figure}

Since the curvature reduction is negligible for pure hard magnetic recording media in combination with the flipped head design, a bilayer structure with $50\%$ hard and $50\%$ soft magnetic material is tested as recording material. In the original paper by Vogler \textit{et al} \cite{vogler_reduction}, a bilayer structure showed significantly higher curvature reduction than the pure hard magnetic recording medium. The total height of the grains is again $h=8\,$nm. The material parameters of the soft magnetic composition can be seen in \Cref{tablesoftmagneticmat}. 
Phase diagrams for the different head designs are calculated and the off-track width is again normalized for comparability reasons. They are shown in \Cref{footprintssm}.  Again, the jitter $\sigma_{\mathrm{down}}$ and $P_{\mathrm{max}}$ are calculated for the footprints and compared in \Cref{tableparametersbilayer}. Due to the higher Curie temperature of the exchange spring recording medium, the down-track jitter is determined at $800\,$K. For the exchange spring recording material the behavior of the curvature is similar to that for pure hard magnetic recording media and seems to be even worse. For the Backshield$_{200}$ head design the decrease of the curvature is only 0.08$\%$ compared to the conventional design. However, the maximum switching probability stays 100$\%$ and the down-track jitter decreases.

\begin{center}
\begin{table*}
\centering
\begin{tabular}{|>{\centering}m{2.2cm}|>{\centering}m{1.7cm}|>{\centering}m{3.0cm}|c|c|}
\hline
Curie temp. $T_{\mathrm{\textbf{C}}}$ (K) & Damping $\alpha$& Anisotropy const. $K_{1}$ (MJ/m$^3$)& $J_{\mathrm{s}}$ (T) \\
    \hline
	740.9 & 0.1&$0$  & 1.35  \\
    \hline
 \end{tabular}
\caption{Material parameters of the soft magnetic composition of the bilayer structure.  }
\label{tablesoftmagneticmat}
\end{table*}
\end{center}

\begin{table*}
\begin{tabular}{|l|c|>{\centering}m{1.8cm}|c|c|r|}\hline
  Geometry
& $c$ ($\%$) \,\,\,
&  $P_{\mathrm{max}}$ ($\%$)  &$\sigma_{\mathrm{down}}$ ($\%$) \\
    \hline
    Basic$_{200}$&$+1.38$&$-2.5$&$-16.2$\\
   Backshield$_{100}$&$+1.9$&$-1.19$&$-4.7$\\
    Backshield$_{200}$&$-0.08$&$+/-0.0$ & $-19$\\
  \hline
 \end{tabular}
\caption{Resulting transition curvature, $P_{\mathrm{max}}$ and jitter parameters of the different flipped head designs in combination with an exchange spring recording medium compared to a conventional recording head design with the same recording material.}
\label{tableparametersbilayer}
\end{table*}

\subsection{Comparison with ERTW model}
To understand why both the pure hard magnetic and the exchange spring media show almost no curvature reduction, the results are compared to theoretical considerations with the effective recording time window (ERTW) model described by Vogler \textit{et al} \cite{basic,vogler_reduction}.
The effective recording time window (ERTW) is defined by \cite{zhu2013understanding,basic}

\begin{align}
    \mathrm{ERTW}_{\uparrow} = [t(T_{\mathrm{c}}),t(T_{\mathrm{freeze}})]\, \cap \, [t_{\uparrow,\mathrm{start}},t_{\uparrow,\mathrm{final}}]
\end{align}

The first term on the right hand side gives the time window in which the grains can be written in HAMR, namely in the temperature range between the Curie temperature $T_{\mathrm{c}}$ and the freezing temperature $T_{\mathrm{freeze}}$. At the freezing temperature the coercivity decreases below the given field strength such that the grain cannot be written any longer. To calculate the ERTW, the freezing temperature has to be estimated. For this reason, hysteresis loops are simulated for various temperatures with VAMPIRE \cite{evans}. At each fixed temperature, 128 hysteresis loops are computed and the temperature dependence of the coercive field d$H_{\mathrm{c}}/$d$T$ is determined. If the coercive field at a fixed temperature is lower than the write field, the temperature is a possible write temperature. Since the magnitude of the write field is not constant for the flipped head designs, a correction factor is included in the model. With the correction factor the magnitude of the field is updated according to the field gradient and the resulting write field is used for the determination of the freezing temperature.\\ The second term gives the time window in which the field points in the desired write direction, which in the following is regarded as pointing upwards without loss of generality.
With this ERTW definition a switching probability phase diagram without noise can be computed for different materials and various fields and field gradients. The switching probability of a recording grain can be computed via \cite{basic}

\begin{align}
    p= \min\left(\frac{\mathrm{ERTW}_{\uparrow}}{\Theta_{\mathrm{ERTW}}},1\right)\left[1-\min\left(\frac{\mathrm{ERTW}_{\downarrow}}{\Theta_{\mathrm{ERTW}}},1\right)\right].
\end{align}

The first term gives the probability that a bit is written in write direction. Since the probability cannot exceed 1, the minimum between 1 and $\Theta_{\mathrm{ERTW}}$ is taken. $\Theta_{\mathrm{ERTW}}$ is a threshold of the ERTW that gives the time that must be exceeded for successful switching to occur. It can be determined as a fit parameter to reproduce the results by the LLB model. The second term is the probability that a grain is overwritten after it was already aligned in write direction. Together, $p$ describes the joint probability to switch a bit in write direction and not reverting it afterwards. From the phase diagrams, the transition curvature is calculated and it is determined how much the theoretical curvature reduction should be.\\
For the Backshield$_{200}$ geometry, the theoretically expected curvature reduction is about 1$\%$ for the pure hard magnetic recording material compared to the conventional design. This agrees very good with the curvature reduction resulting from the LLB simulations. 
For the exchange spring bilayer structure, the ERTW model predicts a curvature reduction of about $1.7\%$. The very small value achieved by the LLB simulation ($-0.08\%$) results from the switch of the transition around zero. This switch does not happen in the analytical ERTW model and comes from the stochastic nature of the LLB model.\\
Additionally, with the ERTW model, it can be analyzed with reasonable computational effort how the transition curvature reduction depends on the field gradient. The resulting curvature reduction values can be seen in \Cref{tablecurvaturereductionforgradients}. They display that the curvature reduces linearly with the field gradient. Moreover, a higher potential for the curvature reduction can be seen for the exchange spring bilayer structure. However, even with field gradients up to 40\,mT/nm, the bilayer structure did not show as high curvature reduction as the exchange spring media used by Vogler \textit{et al}. This can most likely be explained by the different damping constants used in the hard magnetic material. In the work by Vogler \textit{et al}, the damping constant is $\alpha_{\mathrm{HM}}=0.1$ whereas it is $\alpha_{\mathrm{HM}}=0.02$ in this work. This leads to different $dH_{\mathrm{c}}/dT$ gradients and thus to the different curvature behavior. This shows that damping plays a key role for the curvature reduction.

\begin{center}
\begin{table*}
\centering
\begin{tabular}{|>{\centering}m{2.5cm}|c|c|c|c|}\hline
 Field gradient [mT/nm] & $c$ ($\%$) HM &$c$ ($\%$) HM/SM\,\\
  \hline
  	12&-1.0&-1.7\\
   16& -1.6& -3.64\\
    20& -2.09 &-5.58\\
   24& -2.59 &-7.5\\
   28& -3.38 &-9.48\\
   32& -4.6 &-11.4\\
   36& -5.2 &-13.11\\
   40& -5.8 &-14.2\\
  \hline
 \end{tabular}
\caption{Curvature reduction in $\%$ compared to a conventional design for pure hard magnetic recording material (HM) and an exchange spring bilayer structure (HM/SM) calculated with the ERTW model.}
\label{tablecurvaturereductionforgradients}
\end{table*}
\end{center}

\section{Discussion}
To conclude, in this work we tried to optimize the design of the write pole of a recording head for heat-assisted magnetic recording in order to reduce transition curvature. The write pole of the head was optimized in a way to maximize both the $z-$field and the $z-$field gradient at the position where the applied heat pulse is cooling down. This was done with the help of topology optimization which is an application of the inverse magnetostatic problem. Different starting geometries were considered. The comparison of the different geometries to a conventional recording head design showed the best results for a write pole and an additional backshield which both have dimension $200\,$nm in down-track direction. 
The resulting field gradient is then 11.8\,mT/nm. 
The optimized geometries are all similar. They all have smooth edges, a tapered shape with a tip in the middle and the peak is skewed in $x-$direction such that the distance of the pole tip to the recording medium is close to the fieldbox and larger on the side away from it (see \Cref{recordinghead}). Since all optimized geometries are similar, the field and the field gradient could be maximized further by optimizing even larger write poles with an additional backshield. Another option to further increase the field gradient produced by the write pole is a design where the NFT is closer to the write pole.\\
We calculated switching probability phase diagrams of FePt like head magnetic recording media for the different optimized field gradients. It is noteworthy that the switching performance, in terms of down-track jitter and maximum switching probability, for all flipped head designs is worse than that of the conventional head design. Both, the AC and the DC noise increase for the flipped head designs. 
The performance loss results most likely from the smaller write fields that are produced by the flipped head designs.
Analyzing the transition curvature showed that the curvature reduces only marginally even for the Backshield$_{200}$ design where the highest write field and field gradient can be achieved. 
An idea how to improve the curvature reduction is to use an exchange spring medium \cite{suessexchange,wangexchange,victora,suess, coffey, suess1}. The soft magnetic layer acts as a write assist and thus smaller write fields are required to write the grains. For the conventional design the curvature of the exchange spring recording medium is larger than that of the pure hard magnetic medium. Thus, the curvature reduction potential was expected to be higher for the exchange spring media. We computed switching probability phase diagrams for the different head designs in combination with the exchange spring recording media. Surprisingly, the curvature reduction for the Backshield$_{200}$ design was even smaller than for the pure hard magnetic material. 
To understand why the curvature reduces only so little, we analyzed the results in the context of the effective recording time window (ERTW) model that was used in the paper by Vogler \textit{et al} \cite{vogler_reduction}. The ERTW model confirmed the results of the LLB model. Hence, we can conclude that the marginal reduction of the curvature is not a consequence of noise. The analysis also pointed out that the exchange spring recording medium shows a larger potential for curvature reduction if higher field gradients are used but even here the curvature reduction is only marginally.\\
In conclusion, we did not see any improvement of the transition curvature for the resulting field gradients of the optimized flipped head design in combination with pure hard magnetic and exchange spring recording media. The results showed that a simple optimization of the conventional head design design is not sufficient for effective curvature reduction but that new head concepts have to be introduced to reduce transition curvature.

\section{Acknowledgements}
The authors would like to thank the Vienna Science and Technology Fund (WWTF) under grant No. MA14-044, the Advanced Storage Technology Consortium (ASTC), and the Austrian Science Fund (FWF) under grant No. I2214-N20 for financial support. The computational results presented have been achieved using the Vienna Scientific Cluster (VSC).


\begin{thebibliography}{10}

\bibitem{ersteshamr}
Hiroshi Kobayashi, Motoharu Tanaka, Hajime Machida, Takashi Yano, and Uee~Myong
  Hwang.
\newblock {\em Thermomagnetic recording}.
\newblock Google Patents, August 1984.

\bibitem{fan}
C.~Mee and G.~Fan.
\newblock A proposed beam-addressable memory.
\newblock {\em IEEE Transactions on Magnetics}, 3(1):72--76, 1967.

\bibitem{hamr1}
Robert~E. Rottmayer, Sharat Batra, Dorothea Buechel, William~A. Challener,
  Julius Hohlfeld, Yukiko Kubota, Lei Li, Bin Lu, Christophe Mihalcea, Keith
  Mountfield, and {others}.
\newblock Heat-assisted magnetic recording.
\newblock {\em IEEE Transactions on Magnetics}, 42(10):2417--2421, 2006.

\bibitem{hamr}
Mark~H. Kryder, Edward~C. Gage, Terry~W. McDaniel, William~A. Challener,
  Robert~E. Rottmayer, Ganping Ju, Yiao-Tee Hsia, and M.~Fatih Erden.
\newblock Heat assisted magnetic recording.
\newblock {\em Proceedings of the IEEE}, 96(11):1810--1835, 2008.

\bibitem{evans}
R.~F.~L. Evans, Roy~W. Chantrell, Ulrich Nowak, Andreas Lyberatos, and H.-J.
  Richter.
\newblock Thermally induced error: {Density} limit for magnetic data storage.
\newblock {\em Applied Physics Letters}, 100(10):102402, 2012.

\bibitem{nft1}
Nan Zhou, Xianfan Xu, Aaron~T. Hammack, Barry~C. Stipe, Kaizhong Gao, Werner
  Scholz, and Edward~C. Gage.
\newblock Plasmonic near-field transducer for heat-assisted magnetic recording.
\newblock {\em Nanophotonics}, 3(3), January 2014.

\bibitem{nft2}
Jacek Gosciniak, Marcus Mooney, Mark Gubbins, and Brian Corbett.
\newblock Novel droplet near-field transducer for heat-assisted magnetic
  recording.
\newblock {\em Nanophotonics}, 4(1), January 2015.

\bibitem{zhu_correcting}
J.~G.~J. Zhu and H.~Li.
\newblock Correcting {Transition} {Curvature} in {Heat}-{Assisted} {Magnetic}
  {Recording}.
\newblock {\em IEEE Transactions on Magnetics}, 53(2):1--7, February 2017.

\bibitem{zhu_field}
Jian-Gang~(Jimmy) Zhu and Hai Li.
\newblock Write head field design for correcting transition curvature in heat
  assisted magnetic recording.
\newblock {\em AIP Advances}, 7(5):056505, February 2017.

\bibitem{zhuli}
Yuwei Qin, Hai Li, and Jian-Gang Zhu.
\newblock Curvature-{Eliminating} {Head} {Field} and {Track} {Edge}
  {Characteristics} in {Heat}-{Assisted} {Magnetic} {Recording}.
\newblock {\em IEEE Transactions on Magnetics}, 53(11):1--4, 2017.

\bibitem{vogler_reduction}
Christoph Vogler, Claas Abert, Florian Bruckner, and Dieter Suess.
\newblock Efficiently reducing transition curvature in heat-assisted magnetic
  recording with state-of-the-art write heads.
\newblock {\em Applied Physics Letters}, 110(18):182406, May 2017.

\bibitem{suh_head}
Sung-dong Suh, Young-hun Im, and Hae-Sung Kim.
\newblock Heat-assisted magnetic recording head, October 2009.

\bibitem{lee_head}
Myung-bok Lee and Jin-Seung Sohn.
\newblock Heat assisted magnetic recording head, July 2007.

\bibitem{bendsoe_topology_2013}
Martin~Philip Bendsoe and Ole Sigmund.
\newblock {\em Topology {Optimization}: {Theory}, {Methods}, and
  {Applications}}.
\newblock Springer Science \& Business Media, April 2013.
\newblock Google-Books-ID: ZCjsCAAAQBAJ.

\bibitem{bruckner_solving}
Florian Bruckner, Claas Abert, Gregor Wautischer, Christian Huber, Christoph
  Vogler, Michael Hinze, and Dieter Suess.
\newblock Solving {Large}-{Scale} {Inverse} {Magnetostatic} {Problems} using
  the {Adjoint} {Method}.
\newblock {\em Scientific Reports}, 7:40816, January 2017.

\bibitem{LLB}
Christoph Vogler, Claas Abert, Florian Bruckner, and Dieter Suess.
\newblock Landau-{Lifshitz}-{Bloch} equation for exchange-coupled grains.
\newblock {\em Physical Review B}, 90(21):214431, 2014.

\bibitem{huber_topology}
C.~Huber, C.~Abert, F.~Bruckner, C.~Pfaff, J.~Kriwet, M.~Groenefeld,
  I.~Teliban, C.~Vogler, and D.~Suess.
\newblock Topology optimized and 3d printed polymer-bonded permanent magnets
  for a predefined external field.
\newblock {\em Journal of Applied Physics}, 122(5):053904, August 2017.

\bibitem{abert2017fast}
Claas Abert, Christian Huber, Florian Bruckner, Christoph Vogler, Gregor
  Wautischer, and Dieter Suess.
\newblock A fast finite-difference algorithm for topology optimization of
  permanent magnets.
\newblock {\em Journal of Applied Physics}, 122(11):113904, 2017.

\bibitem{choi}
Jae~Seok Choi and Jeonghoon Yoo.
\newblock Simultaneous structural topology optimization of electromagnetic
  sources and ferromagnetic materials.
\newblock {\em Computer Methods in Applied Mechanics and Engineering},
  198(27-29):2111--2121, May 2009.

\bibitem{huberthesis}
Christian Huber.
\newblock {\em 3D printed polymer-bonded NdFeB magnets for a tailored magnetic
  field}.
\newblock PhD thesis, TU Wien, 2017.

\bibitem{fenics1}
Todd Dupont, Johan Hoffman, Claus Johnson, Robert~C. Kirby, Mats~G. Larson,
  Anders Logg, and L.~Ridgway Scott.
\newblock {\em The fenics project}.
\newblock Chalmers Finite Element Centre, Chalmers University of Technology,
  2003.

\bibitem{fenics2}
Martin~S. Alna~es, Jan Blechta, Johan Hake, August Johansson, Benjamin Kehlet,
  Anders Logg, Chris Richardson, Johannes Ring, Marie~E. Rognes, and Garth~N.
  Wells.
\newblock The {FEniCS} project version 1.5.
\newblock {\em Archive of Numerical Software}, 3(100):9--23, 2015.

\bibitem{dolfin1}
S~W Funke and P~E Farrell.
\newblock A framework for automated {PDE}-constrained optimisation.
\newblock {\em ACM Transactions on Mathematical Software}, page~28.

\bibitem{dolfin2}
P.~E. Farrell, D.~A. Ham, S.~W. Funke, and M.~E. Rognes.
\newblock Automated {Derivation} of the {Adjoint} of {High}-{Level} {Transient}
  {Finite} {Element} {Programs}.
\newblock {\em SIAM Journal on Scientific Computing}, 35(4):C369--C393, January
  2013.

\bibitem{lbfgsb1}
Richard~H. Byrd, Peihuang Lu, Jorge Nocedal, and Ciyou Zhu.
\newblock A limited memory algorithm for bound constrained optimization.
\newblock {\em SIAM Journal on Scientific Computing}, 16(5):1190--1208, 1995.

\bibitem{lbfgsb2}
Ciyou Zhu, Richard~H. Byrd, Peihuang Lu, and Jorge Nocedal.
\newblock Algorithm 778: {L}-{BFGS}-{B}: {Fortran} subroutines for large-scale
  bound-constrained optimization.
\newblock {\em ACM Transactions on Mathematical Software (TOMS)},
  23(4):550--560, 1997.

\bibitem{recording_head}
Michael~Allen Seigler, Mark~William Covington, Michael~Leigh Mallary, Hua Zhou,
  and Amit~Vasant Itagi.
\newblock Recording head for heat assisted magnetic recording, May 2013.

\bibitem{brug_magnetic_1996}
James~A. Brug, Thomas~C. Anthony, and Janice~H. Nickel.
\newblock Magnetic {Recording} {Head} {Materials}.
\newblock {\em MRS Bulletin}, 21(09):23--27, September 1996.

\bibitem{kief_materials}
M.T. Kief and R.H. Victora.
\newblock Materials for heat-assisted magnetic recording.
\newblock {\em MRS Bulletin}, 43(02):87--92, February 2018.

\bibitem{feco}
Y.~Okada, H.~Hoshiya, T.~Okada, and M.~Fuyama.
\newblock Magnetic properties of {FeCo} multilayered films for single pole
  heads.
\newblock {\em IEEE Transactions on Magnetics}, 40(4):2368--2370, July 2004.

\bibitem{hmsmedia}
J.~D. Kiely, P.~M. Jones, H.~Wang, R.~Yang, W.~Scholz, M.~Benakli, J.~L. Brand,
  and S.~Gangopadhyay.
\newblock Media {Roughness} and {Head}-{Media} {Spacing} in {Heat}-{Assisted}
  {Magnetic} {Recording}.
\newblock {\em IEEE Transactions on Magnetics}, 50(3):132--136, March 2014.

\bibitem{abert_magnum.fe:_2013}
Claas Abert, Lukas Exl, Florian Bruckner, André Drews, and Dieter Suess.
\newblock magnum.fe: {A} micromagnetic finite-element simulation code based on
  {FEniCS}.
\newblock {\em Journal of Magnetism and Magnetic Materials}, 345:29--35,
  November 2013.

\bibitem{hamann2007heating}
Hendrik~F Hamann, Prakash Kasiraj, Jeffrey~S Lille, Yves~C Martin, Chie~Ching
  Poon, Neil~Leslie Robertson, Jan-Ulrich Thiele, and Hemantha~Kumar
  Wickramasinghe.
\newblock Heating device and magnetic recording head for thermally-assisted
  recording, August~28 2007.
\newblock US Patent 7,262,936.

\bibitem{huang2014heat}
Pin-Wei Huang and Randall~H Victora.
\newblock Heat assisted magnetic recording: Grain size dependency, enhanced
  damping, and a simulation/experiment comparison.
\newblock {\em Journal of Applied Physics}, 115(17):17B710, 2014.

\bibitem{zhu2013understanding}
Jian-Gang Zhu and Hai Li.
\newblock Understanding signal and noise in heat assisted magnetic recording.
\newblock {\em IEEE Transactions on Magnetics}, 49(2):765--772, 2013.

\bibitem{basic}
Christoph Vogler, Claas Abert, Florian Bruckner, Dieter Suess, and Dirk
  Praetorius.
\newblock Basic noise mechanisms of heat-assisted-magnetic recording.
\newblock {\em Journal of Applied Physics}, 120(15):153901, 2016.

\bibitem{suessexchange}
Dieter Suess, Thomas Schrefl, S.~Fähler, Markus Kirschner, Gino Hrkac, Florian
  Dorfbauer, and Josef Fidler.
\newblock Exchange spring media for perpendicular recording.
\newblock {\em Applied Physics Letters}, 87(1):012504, 2005.

\bibitem{wangexchange}
Jian-Ping Wang, Weikang Shen, and Jianmin Bai.
\newblock Exchange coupled composite media for perpendicular magnetic
  recording.
\newblock {\em IEEE transactions on magnetics}, 41(10):3181--3186, 2005.

\bibitem{victora}
R.~H. Victora and X.~Shen.
\newblock Exchange coupled composite media for perpendicular magnetic
  recording.
\newblock {\em IEEE Transactions on Magnetics}, 41(10):2828--2833, October
  2005.

\bibitem{suess}
Dieter Suess.
\newblock Micromagnetics of exchange spring media: {Optimization} and limits.
\newblock {\em Journal of magnetism and magnetic materials}, 308(2):183--197,
  2007.

\bibitem{coffey}
Kevin~Robert Coffey, Jan-Ulrich Thiele, and Dieter~Klaus Weller.
\newblock {\em ‘{Thermal} spring’magnetic recording media for writing using
  magnetic and thermal gradients}.
\newblock Google Patents, April 2005.

\bibitem{suess1}
Dieter Suess and Thomas Schrefl.
\newblock Breaking the thermally induced write error in heat assisted recording
  by using low and high {Tc} materials.
\newblock {\em Applied Physics Letters}, 102(16):162405, 2013.

\bibitem{nft3}
W.~A. Challener, Chubing Peng, A.~V. Itagi, D.~Karns, Wei Peng, Yingguo Peng,
  XiaoMin Yang, Xiaobin Zhu, N.~J. Gokemeijer, Y.-T. Hsia, G.~Ju, Robert~E.
  Rottmayer, Michael~A. Seigler, and E.~C. Gage.
\newblock Heat-assisted magnetic recording by a near-field transducer with
  efficient optical energy transfer.
\newblock {\em Nature Photonics}, 3(4):220--224, April 2009.

\bibitem{curvatureanalysis}
M.~Hashimoto, M.~Salo, Y.~Ikeda, A.~Moser, R.~Wood, and H.~Muraoka.
\newblock Analysis of written transition curvature in perpendicular magnetic
  recording from spin-stand testing.
\newblock {\em IEEE transactions on magnetics}, 43(7):3315--3319, 2007.

\end{thebibliography}

\end{document}